\documentclass[aps,prc,nofootinbib,showkeys,showpacs,superscriptaddress,preprint,onecolumn]{revtex4-1}

\usepackage{graphicx}
\usepackage{dcolumn}
\usepackage{bm}
\usepackage{amsmath,amsfonts,amsthm,amssymb}
\usepackage{hyperref} 
\usepackage{float}
\usepackage{lipsum}
\usepackage{physics}
\usepackage{nicefrac}
\usepackage{yfonts}
\usepackage[utf8]{inputenc}
\usepackage{cancel}

\usepackage[dvipsnames]{xcolor}

\usepackage[sort&compress]{natbib}
\hypersetup{
colorlinks = true,
linkcolor = blue,
anchorcolor = blue,
citecolor = blue,
filecolor = blue,
urlcolor = blue
}
\usepackage[normalem]{ulem}

\begin{document}

\title{Hyperon longitudinal polarization and vector meson spin alignment in a thermal model for heavy-ion collisions}

\author{Soham Banerjee}
\email{soham.banerjee@niser.ac.in}
\affiliation{School of Physical Sciences, National Institute of Science Education and Research, An OCC of Homi Bhabha National Institute, Jatni, 752050, India}
            
\author{Samapan Bhadury}
\email{samapanb@iiserbpr.ac.in}
\affiliation{Department of Physical Sciences, Indian Institute of Science Education and Research Berhampur, Laudigam-760003, District-Ganjam, Odisha, India}
            
\author{Wojciech Florkowski}
\email{wojciech.florkowski@uj.edu.pl}
\affiliation{Institute of Theoretical Physics, Jagiellonian University, Kraków, 30-348, Poland}   

\author{Amaresh Jaiswal}
\email{a.jaiswal@niser.ac.in}
\affiliation{School of Physical Sciences, National Institute of Science Education and Research, An OCC of Homi Bhabha National Institute, Jatni, 752050, India}

\author{Radoslaw Ryblewski}
\email{radoslaw.ryblewski@ifj.edu.pl}
\affiliation{Institute of Nuclear Physics, Polish Academy of Sciences, Kraków, 31-342, Poland}  

\begin{abstract}
The concept of a common local spin equilibrium for both spin-1/2 and spin-1 particles is incorporated into a thermal model of particle production in heavy-ion collisions at the top RHIC energies. We show that an effective spin polarization tensor leading to a correct description of the longitudinal spin polarization of $\Lambda$ hyperons simultaneously yields a positive alignment of vector mesons ($\phi$ and $K^{*0}$) that grows monotonically with transverse momentum and centrality. Similar trends can be seen in the data, suggesting a possible common mechanism for longitudinal spin polarization and alignment. However, model calculations are insufficient to explain the data in a fully quantitative way. The correlation found between the magnitude of the $\Lambda$ longitudinal polarization and vector meson alignment suggests further more elaborate investigations of this issue.

\end{abstract}

\maketitle

\section{Introduction}
\label{introduction}

The experimental investigations of spin polarization effects in relativistic heavy-ion collisions offer new possibilities for studying the properties of strongly interacting matter. These studies focus mainly on measurements of hyperon spin polarization~\cite{STAR:2017ckg, STAR:2018gyt, STAR:2019erd,ALICE:2019onw,STAR:2020xbm,ALICE:2021pzu,STAR:2021beb,HADES:2022enx,STAR:2023nvo} and vector-meson spin alignment~\cite{ALICE:2019aid,STAR:2022fan,ALICE:2022dyy}; for recent reviews, see~\cite{Becattini:2024uha,Niida:2024ntm,Chen:2024afy} and for various related theoretical developments, see~\cite{Liang:2004ph, Liang:2004xn, Becattini:2009wh, Yang:2017sdk, Montenegro:2017rbu, Montenegro:2017lvf, Florkowski:2017ruc, Florkowski:2017dyn, Florkowski:2018ahw, Sun:2018bjl, Sheng:2019kmk, Hattori:2019lfp, Xie:2019jun, Florkowski:2019qdp, Weickgenannt:2019dks, Weickgenannt:2020aaf, Bhadury:2020cop,Shi:2020htn, Singh:2020rht,Yi:2021ryh, Wang:2021ngp, Hongo:2021ona, Florkowski:2021wvk, Wang:2021wqq, Sheng:2022wsy,Sheng:2022ffb,Hongo:2022izs,Weickgenannt:2022zxs, Sarwar:2022yzs, Wagner:2022amr,Wagner:2022gza, Kumar:2022ylt, Biswas:2023qsw, Weickgenannt:2023btk, Kiamari:2023fbe,Fang:2023bbw,Palermo:2023cup, Palermo:2023qfg,Kumar:2023ojl, Kumar:2023ghs,Chen:2023hnb,Goncalves:2024xzo,Grossi:2024pyh,Sun:2024anu, Grossi:2025hxg, Arslan:2025tan}.

For some time, the main difficulty in describing the $\Lambda$ polarization data was to reproduce the experimental results for the longitudinal component of the polarization vector (i.e., the component along the beam direction)~\cite{Becattini:2017gcx}. The proposed solutions to this problem include now adding the effects due to thermal shear~\cite{Liu:2021uhn,Fu:2021pok, Becattini:2021suc, Becattini:2021iol} and/or a variation of the mass of the spin-polarized particle~\cite{Fu:2021pok}. Yet another solution is to use the spin polarization tensor obtained from spin hydrodynamics~\cite{Singh:2024cub, Sapna:2025yss}. On the other hand, many theoretical frameworks yield a negative spin alignment of vector mesons~\cite{Liang:2004xn, Yang:2017sdk, Sheng:2019kmk, Kumar:2023ghs, Sheng:2022wsy}, which is incompatible with the data, with the exception of the model presented in \cite{Sheng:2019kmk, Grossi:2025hxg, Liu:2025klr}.

In this work, we use a recently proposed uniform description for the local equilibrium of spin-1/2 and spin-1 particles~\cite{Florkowski:2026ofs} to simultaneously address the two problems outlined above. To make our study  as simple as possible, we use the framework of the thermal model used before to analyze the $\Lambda$ polarization at the top RHIC energies~\cite{Broniowski:2002wp, Florkowski:2019voj,Florkowski:2021xvy,Banerjee:2024xnd}. Since we consider high energies, the model is boost-invariant, however, it incorporates an anisotropic form of the transverse-flow responsible for the non-zero elliptic flow.

The conclusion of~\cite{Banerjee:2024xnd} is that the longitudinal spin polarization can be described by using a thermal vorticity, $\varpi_{\mu \nu}=-\frac{1}{2}\left(\partial_\mu \beta_\nu-\partial_\nu \beta_\mu\right)$ (where $\beta_\mu$ is the ratio of fluid four-velocity $u_\mu$ to temperature $T$) with neglected ``electric''-like components. As found in our previous work~\cite{Florkowski:2019voj}, this procedure is consistent with a non-relativistic treatment where the polarization is solely determined by the spatial components of the rotation $\partial_i v_j - \partial_j v_i$ \cite{Voloshin_2018}.  In this work, we additionally introduce a scale parameter $\lambda$ that multiplies the thermal vorticity and is used to check the sensitivity of our results to the magnitude of the longitudinal polarization of $\Lambda$ hyperons. 

With the form of the spin polarization tensor defined as the thermal vorticity (calculated directly in the thermal model) with neglected electric-like part and rescaled by a factor $\lambda$, we calculate the parameter $\rho_{00}$ of the spin density matrix of vector mesons as a function of transverse momentum and centrality. We also present the results for the $p_T$-integrated values and their dependence on the direction of the quantization axis. 

Our results establish a quantitative connection between the spin observables of the spin-1/2 and spin-1 hadrons that are in a common local spin equilibrium. We demonstrate that a non-trivial spin alignment arises in such a state without requiring the inclusion of dissipative phenomena -- as the spin alignment is usually attributed to the spin fluctuations, it may appear also in the state of local equilibrium where only the mean values are controlled by the Lagrange multipliers.

The positivity and growth of the alignment with transverse-momentum and centrality found in our model agrees with general trends observed in the data; however, a satisfactory quantitative description of the data is not reproduced. This suggests that more realistic calculations of the spin polarization tensor should be used in the near future to analyze the data. This is also indicated by the dependence of our results on the parameter $\lambda$ -- its growth from 1 to 3 improves the description of both the longitudinal polarization and alignment. This may be a signal for a deeper connection between these two effects that is worth pursuing. 

The paper is organized as follows. In Sec.~\ref{sec:thermal}, we briefly review the thermal model with anisotropic transverse flow and introduce the projected thermal vorticity together with the scale parameter $\lambda$. Section~\ref{sec:framework} presents the theoretical framework for the spin alignment of vector mesons. Our numerical results for the $p_T$-integrated alignment, its transverse-momentum, centrality,  and quantization-axis dependence are discussed in Sec.~\ref{sec:results}. Finally, Sec.~\ref{sec:summary} contains a summary and outlook.

\smallskip
{\it Notation and conventions}: For the Levi-Civita tensor
$\epsilon^{\mu\nu\alpha\beta}$ we follow the convention $\epsilon^{0123} =-\epsilon_{0123} = +1$. The metric tensor is of the form $g_{\mu\nu} = \textrm{diag}(+1,-1,-1,-1)$. We use boldface symbols to denote three vectors and their scalar products are written as $\bm{a}\cdot\bm{b}$. The particles are always considered on the mass shell with the energy $E_{\bm{p}} = \sqrt{m^2+\bm{p}^2}$. Throughout the text, we make use of natural units, $\hbar = c = k_{\rm B} = 1$.

\section{Thermal Model}
\label{sec:thermal}
We employ a thermal model based on the assumption of simultaneous kinetic and chemical freeze-out~\cite{Broniowski:2001we,Florkowski:2001fp}. In this single freeze-out approach all hadrons decouple from the medium at the same proper time $\tau_{\rm f}$ and temperature $T_{\rm f}$. Despite its simplicity, this model has been remarkably successful in reproducing transverse-momentum spectra and elliptic flow coefficients across a range of centralities in heavy-ion collisions ~\cite{Broniowski:2002wp,Baran:2004kra}.

In its standard formulation, the model is characterized by four parameters: the freeze-out temperature $T_{\rm f}$, the baryon chemical potential $\mu_{\rm B}$, the proper time $\tau_{\rm f}$, and the maximum transverse size of the fireball $r_{\rm max}$. The temperature and the chemical potential are fixed by measuring hadronic yield ratios, while the system size and proper time are extracted from the fits to the transverse momentum spectra. At the ultrarelativistic energies that we are considering here, the net baryon at midrapidity is negligible and we set $\mu_{\rm B}=0$.

\begin{table}[t] 
\centering
\begin{tabular}{|c|c|c|c|c|} 
 \hline
 Centrality\,(\text{\%})  & $\epsilon$ & $\delta$ & $\tau_{\rm f}$ [fm] & $r_{\textrm{max}}$\,[fm] \\ [0.5ex] 
 \hline
  0--15 &~ 0.055~ &~ 0.12~ &~ 7.666~ & ~6.540 \\ [0.5ex] 
 \hline                                
 15--30 &~ 0.097~ &~ 0.26~ &~ 6.258~ & ~5.417\\ [0.5ex] 
 \hline                                
 30--60 &~ 0.137~ &~ 0.37~ &~ 4.266~ & ~3.779 \\ [0.5ex] 
 \hline
\end{tabular}
\caption{Thermal model parameters used to describe the PHENIX data at $\sqrt{s_{\rm NN}}=130~{\rm GeV}$, see Ref.~\cite{Broniowski:2002wp, Baran:2004kra}.}
\label{tab.par}
\end{table}

Since the spin alignment is interconnected to longitudinal spin polarization, which in turn is connected to the anisotropy of collective flow; we make use of an extended version of the thermal model that includes azimuthal asymmetry in the transverse hydrodynamic flow. The four-velocity field is parameterized as 

\begin{equation}
    u^\mu= \frac{1}{N} \left(\,t,\,x\sqrt{1+\delta}\,,
\,y\sqrt{1-\delta}\,,z\,\right),
\end{equation}
where the parameter $N$ is determined from the normalization condition $u^\mu u_\mu = 1$,
\begin{equation}
    N= \sqrt{\tau^2_{\rm f}-(x^2-y^2)\delta},
\end{equation}
and $\tau_{\rm f}$ is the proper time at freeze-out 
\begin{equation}
\tau_{\rm f}^2=t^2-(x^2+y^2+z^2).
\end{equation}

The parameter $\delta$ controls the anisotropy of the transverse flow (the elliptic flow). Together with $\delta$, the parameter $\epsilon$ is introduced that determines the asymmetry in the transverse space. Both $\delta$ and $\epsilon$ were previously fitted to the PHENIX data at $\sqrt{s_{\rm NN}} = 130$~GeV for three centrality classes: $0$--$15\%$, $15$--$30\%$, and $30$--$60\%$, with the freeze-out temperature $T_{\rm f} = 0.165$~GeV~\cite{Broniowski:2002wp,Baran:2004kra}~\footnote{The parameters of the thermal model were fixed for the beam energy $\sqrt{s_{\rm NN}}=130$ GeV, while the polarization data analyzed here refer to $\sqrt{s_{\rm NN}}=200$ GeV. We neglect small differences of the thermal parameters for these two energies.}.

\begin{figure}[t]
  \centering
  \includegraphics[width=0.6\textwidth]{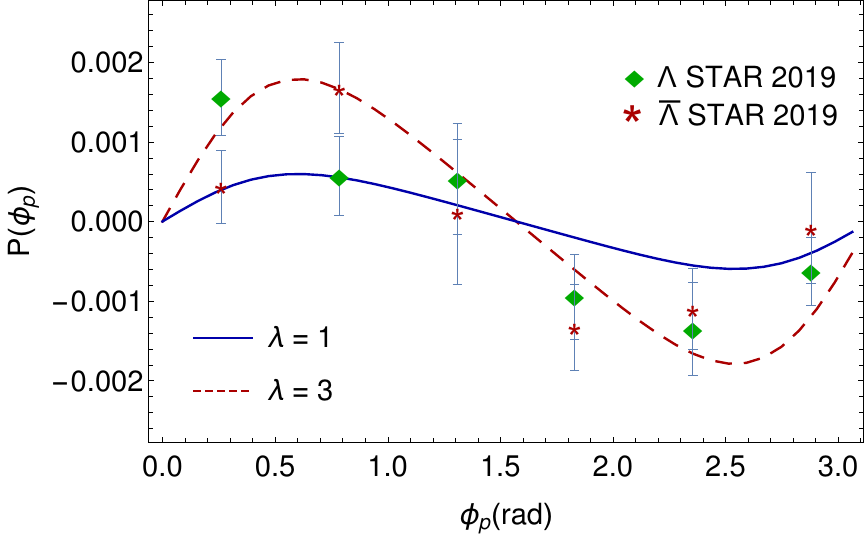}
  \caption{Longitudinal spin polarization of $\Lambda$ hyperons compared to the experimental data \cite{STAR:2019erd}. The blue solid (red dashed) line describes the model result with projected thermal vorticity and $\lambda =1$ ($\lambda =3$).}
  \label{fig:long}
\end{figure}

In our case, we also introduce additional assumptions related to the spin physics: first, we use projected thermal vorticity (the thermal vorticity tensor with all ``electric''-like components neglected, i.e., we set $\varpi_{0i} = 0$), second, we rescale the thermal vorticity by a factor $\lambda$, which controls the magnitude of the longitudinal polarization. We have found that the values $\lambda=1$ and $\lambda=3$ describe the longitudinal polarization of $\Lambda$-hyperons with $\chi^2$ (per one degree of freedom) below 2, see Fig.~\ref{fig:long}.

We note that the thermal model~\cite{Broniowski:2002wp,Baran:2004kra} was not only used by us in previous studies of longitudinal polarization but also used in alignment analyses~\cite{Kumar:2023ojl,Goncalves:2024xzo}, however, in different contexts (influence of second-order hydrodynamic gradients~\cite{Kumar:2023ojl} and spin-vorticity nonequilibrium effects \cite{Goncalves:2024xzo}).

\section{Theoretical framework for spin alignment}
\label{sec:framework}

The spin density matrix for spin-1 particles in the adjoint representation reads~\cite{Florkowski:2026ofs}
\begin{equation}\label{eq:rho_S}
\rho^{S}_{rs*} = \frac{1}{3}\left(\delta_{rs} - \frac{3i}{2} P^k_* \epsilon_{krs} - \sqrt{6}\, T_{rs*}\right),
\end{equation}
where $P^k_*$ is the spin polarization vector and $T_{rs*}$ are the tensor polarizabilities, both defined in the particle rest frame (PRF) denoted by asterisk. Here, the spin densities $r,s$ range from 1 to 3. In local equilibrium, the spin density matrix takes the form
\begin{equation}\label{eq:rho_eq}
\rho^{S}_{\rm eq} = \frac{\exp[\bm{\alpha} \cdot \bm{S}]}{2\cosh\alpha + 1},
\end{equation}
where $\bm{\alpha} = -\bm{b}_*$, with $\bm{b}_*$ being the magnetic-like component of the spin polarization tensor $\omega_{\mu\nu}$ in PRF~\cite{Florkowski:2017dyn}
\begin{equation}\label{eq:bstar}
\bm{b}_* = \frac{1}{m}\left(E_{\bm{p}}\,\bm{b} - \bm{p}\times\bm{e} - \frac{\bm{p}\cdot\bm{b}}{E_{\bm{p}} + m}\,\bm{p}\right).
\end{equation}
The use of projected thermal vorticity means that we use the condition $\bm{e} = 0$. The vector $\bm{S}$ represents three spin matrices in the adjoint representation, $(S^j)_{k\ell} = -i \epsilon_{jk\ell}$. The components of $\bm{b}$ read:
\begin{eqnarray}
b^1 &=&   \frac{\lambda y z }{TN^3}\left(\sqrt{1-\delta} -1 +\delta \right),  \\
b^2 &=&   \frac{\lambda x z}{TN^3}\left( 1+ \delta - \sqrt{1+\delta} \right),   \\
b^3 &=& - \frac{\lambda x y \sqrt{1-\delta^2} }{TN^3}\left( \sqrt{1+\delta} - \sqrt{1-\delta} \right).   
\end{eqnarray}

The alignment $A$ measured experimentally is defined by the central element $\rho_{00}$ of the spin density matrix in the representation where the spin matrix $J_y$ is diagonal~\cite{STAR:2022fan}. Including the effect of a unitary transformation from this representation to the adjoint representation, we find~\cite{Florkowski:2026ofs}
\begin{equation}\label{eq:alignment}
A \equiv \rho_{00} - \frac{1}{3} = -\sqrt{\frac{2}{3}} T^*_{22}
= -\frac{(3\hat{\alpha}_2^2 - 1)(\cosh\alpha - 1)}{3(2\cosh\alpha + 1)},
\end{equation}
where $\hat{\alpha}_2 = \alpha_2/|\bm{\alpha}|$ is the $y$-component of the unit vector along $\bm{\alpha}$. More generally, if the alignment measurement uses the $i$th axis for spin quantization, then the definition of the alignment is
\begin{equation}
A_{i} = -\frac{(3\hat{\alpha}_i^2 - 1)(\cosh\alpha - 1)}{3(2\cosh\alpha + 1)}.
\label{eq:Ai}
\end{equation}
We note that $A_i$'s satisfy the condition $A_{1} + A_{2} + A_{3} = 0$ (which reflects the condition for having a traceless tensor $T^*_{rs}$). We also observe that for an isotropic spin system we have $\hat{\alpha}_i^2 = 1/3 \, (i=1,2,3)$ that yields a vanishing alignment.

The momentum-dependent alignment is obtained by averaging over the freeze-out hypersurface $\Sigma_\mu$:
\begin{equation}\label{eq:Ap}
A(p) = \frac{\displaystyle\int d\Sigma_\mu\, p^\mu\, \mathcal{Z}\, f_0(x,p)\, A(x,p)}{\displaystyle\int d\Sigma_\mu\, p^\mu\, \mathcal{Z}\, f_0(x,p)},
\end{equation}
where $f_0 = \exp[-p\cdot u/T_{\rm f}]$ is the Boltzmann distribution of particles (in our case the $\phi$ and $K^{*0}$ mesons), and $\mathcal{Z} = 2\cosh\alpha + 1$, which is the normalization of spin density matrix.

\begin{figure}[tbp]
  \centering
  \includegraphics[width=0.6\textwidth]{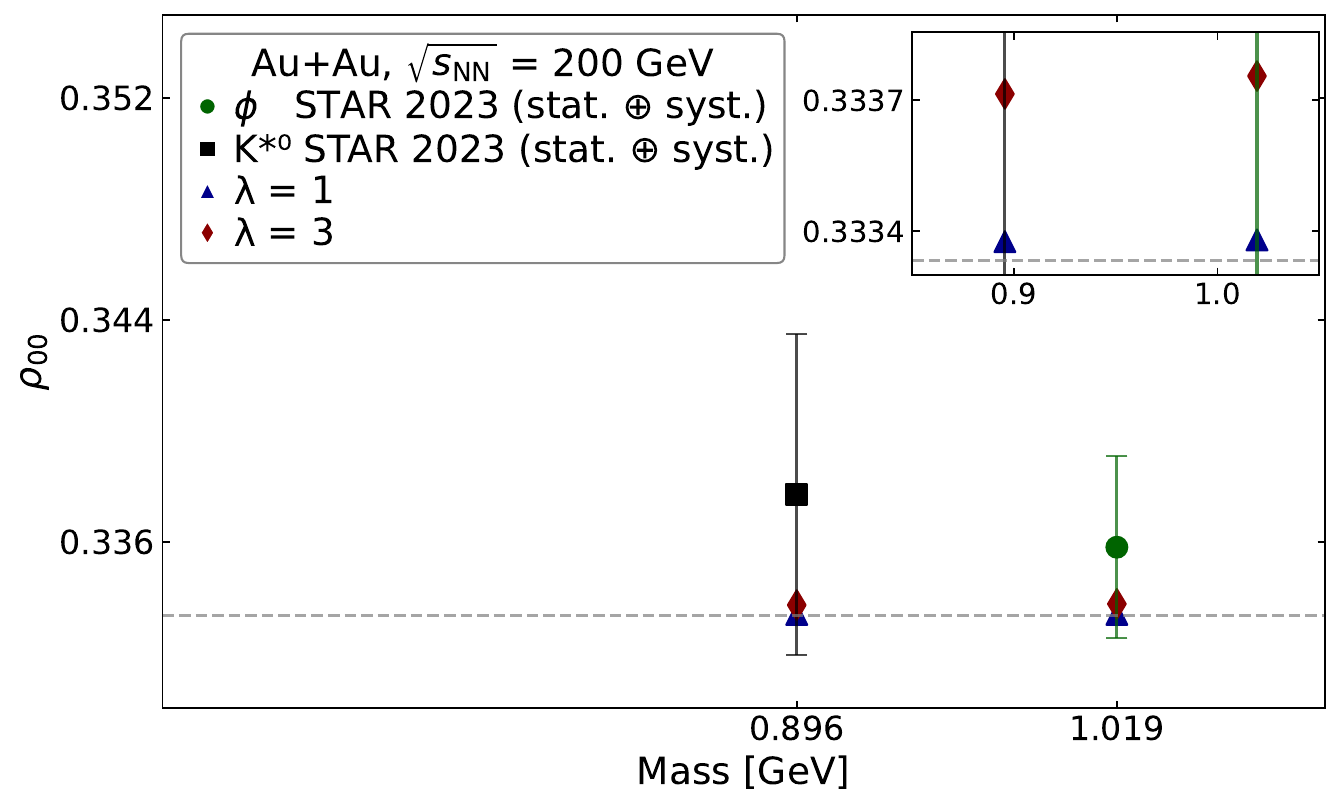}
  \caption{Transverse-momentum integrated $\rho_{00}$ for $\phi$
and $K^{*0}$ mesons as a function of mass. Experimental data from STAR~\cite{STAR:2022fan}.}
  \label{fig:mass_diff_v2}
\end{figure}

\section{Results}
\label{sec:results}
In this section, we present our numerical results for the parameter $\rho_{00}$ of $\phi$ and $K^{*0}$ mesons~\footnote{To compare our results directly with the data we show our results for $\rho_{00}$ rather than for $A$.}. The model parameters used in the calculations are listed in Table~\ref{tab.par}. The masses of the mesons are: $m_\phi = 1.019$~GeV and  $m_{K^{\ast 0}} = 0.892$~GeV.

\subsection{Transverse-momentum integrated alignment}

We begin by examining the $p_T$-integrated values of $\rho_{00}$ for the $\phi$ and $K^{*0}$ mesons. The integration ranges follow the STAR  coverage: $1.2 \leq p_T \leq 5.4$~GeV for $\phi$ mesons and $1.0 \leq p_T \leq 5.0$~GeV for $K^{*0}$ mesons~\cite{STAR:2022fan}. Figure~\ref{fig:mass_diff_v2} displays the $p_T$-integrated $\rho_{00}$ as a function of meson mass for 30--60\% centrality class. Both $\phi$ and $K^{*0}$ mesons exhibit $\rho_{00} > 1/3$, indicating a positive spin alignment. The deviation from $1/3$ is larger for the heavier $\phi$ meson compared to the lighter $K^{*0}$, showing a mass ordering of the alignment. Increasing $\lambda$ from 1 to 3 increases the alignment but does not fully bridge the gap between the model calculations and the data.

\subsection{Transverse-momentum dependence}

\begin{figure}[htbp]
  \centering
  \includegraphics[width=0.6\textwidth]{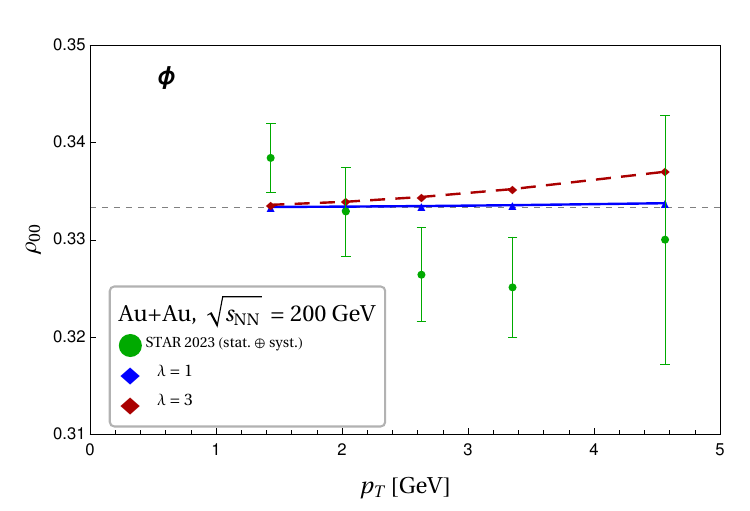}
  \caption{Transverse-momentum dependence of $\rho_{00}$ for $\phi$ mesons. Solid blue and dashed red lines show our results for $\lambda = 1$ and $3$, respectively. Experimental data from STAR~\cite{STAR:2022fan}.}
  \label{fig:phi_diff}
\end{figure}

\begin{figure}[htbp]
  \centering
  \includegraphics[width=0.6\textwidth]{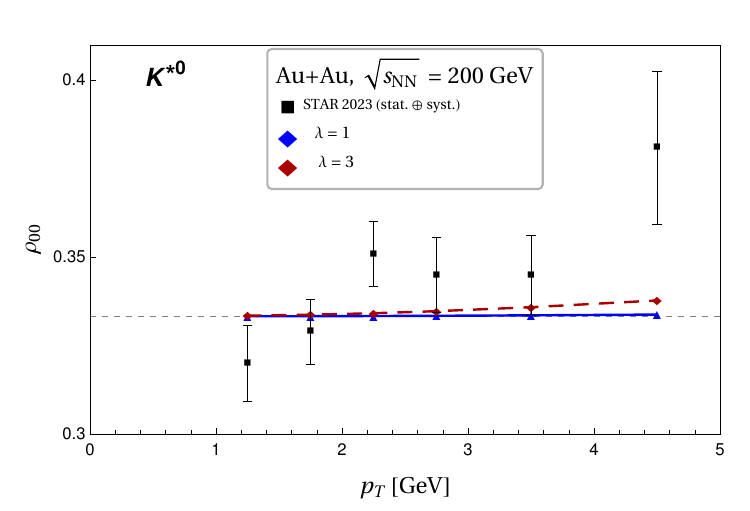}
  \caption{Same as Fig.~\ref{fig:phi_diff} but for $K^{*0}$ mesons.}
  \label{fig:k_diff}
\end{figure}

Figures~\ref{fig:phi_diff} and \ref{fig:k_diff} show the $p_T$-differential $\rho_{00}$ for $\phi$ and $K^{*0}$ mesons, respectively, compared with the STAR data~\cite{STAR:2022fan}. The corresponding numerical values are tabulated in Tables~\ref{tab-phi-pt} and ~\ref{tab-k-pt}. For both species of mesons, the model predicts a monotonically increasing $\rho_{00}$ with $p_T$. The case $\lambda = 3$ produces deviations at the level $10^{-3}$ (for the largest values of momenta).

\begin{table}[htbp]
\centering
\caption{Theoretical values of $\rho_{00}$ for $\phi$ mesons with $\lambda = 1$ and $\lambda = 3$ compared to STAR measurements~\cite{STAR:2022fan} with respect to the second-order event plane. }
\vspace{1.5mm}
\begin{tabular}{|c|cc|c|}
\hline
$p_T$ & \multicolumn{2}{c|}{$\rho_{00}^{\text{th}}$} & $\rho_{00}$ \\
\cline{2-3}
[GeV] & $\lambda=1$ & $\lambda=3$ &  \\
\hline
~1.4339~ &~ 0.333361~ & 0.333581~ &~ 0.3384~  \\
~2.0306~ &~ 0.333396~ & 0.333897~ &~ 0.3329~  \\
~2.6304~ &~ 0.333451~ & 0.334388~ &~ 0.3264~  \\
~3.3535~ &~ 0.333543~ & 0.335199~ &~ 0.3251~  \\
~4.5640~ &~ 0.333747~ & 0.336982~ &~ 0.3300~  \\
\hline
\end{tabular}
\label{tab-phi-pt}
\end{table}

\begin{table}[htbp]
\centering
\caption{Same as Table~\ref{tab-phi-pt} but for $K^{*0}$ mesons.}
\vspace{1.5mm}
\begin{tabular}{|c|cc|c|}
\hline
$p_T$ & \multicolumn{2}{c|}{$\rho_{00}^{\text{th}}$} & $\rho_{00}$ \\
\cline{2-3}
[GeV] & $\lambda=1$ & $\lambda=3$ &  \\
\hline
 ~1.25 ~ &~ 0.333356~ & 0.333538~ &~ 0.320~     \\
 ~1.75 ~ &~ 0.333386~ & 0.333804~ &~ 0.329~     \\
 ~2.25 ~ &~ 0.333433~ & 0.334222~ &~ 0.351~     \\
 ~2.75 ~ &~ 0.333498~ & 0.334797~ &~ 0.345~     \\
 ~3.50 ~ &~ 0.333626~ & 0.335928~ &~ 0.345~     \\
 ~4.50 ~ &~ 0.333845~ & 0.337830~ &~ 0.381~     \\
\hline
\end{tabular}
\label{tab-k-pt}
\end{table}

\subsection{Centrality dependence}
\begin{figure}[htbp]
  \centering
  \includegraphics[width=0.6\textwidth]{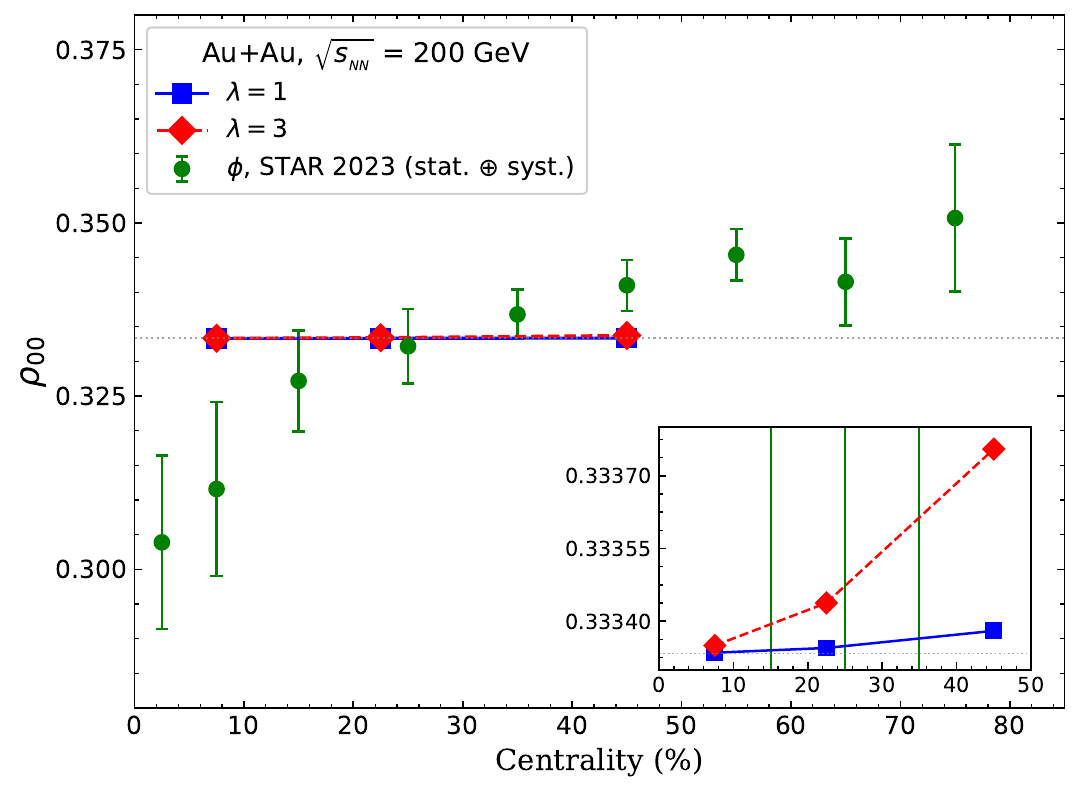}
  \caption{Centrality dependence of $\rho_{00}$ for $\phi$ mesons. Solid blue and dashed red lines show model results for $\lambda = 1$ and $3$, respectively. Experimental data are from STAR~\cite{STAR:2022fan}.}
  \label{fig:centrality_phi}
\end{figure}

\begin{figure}[htbp]
  \centering
  \includegraphics[width=0.6\textwidth]{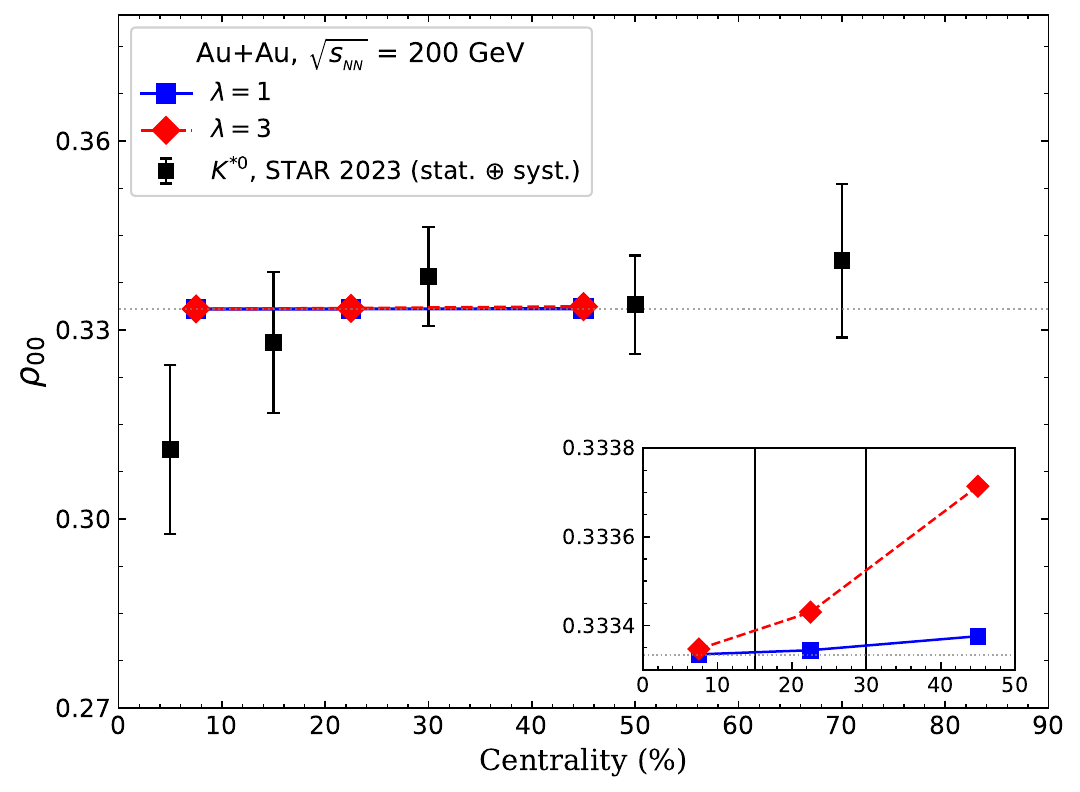}
  \caption{Same as Fig.~\ref{fig:centrality_phi} but for $K^{*0}$ mesons.}
  \label{fig:centrality_k}
\end{figure}

Figures~\ref{fig:centrality_phi} and \ref{fig:centrality_k} present the centrality dependence of $\rho_{00}$ for $\phi$ and $K^{*0}$ mesons, respectively. The corresponding numerical values are collected in Table~\ref{tab-phi-cent}. The model predicts an  increase of $\rho_{00}$ from central to peripheral collisions, however, the centrality dependence observed in the experiment is much stronger.

\begin{table}[htbp]
\centering
\caption{Centrality dependence of theoretical $\rho_{00}$ for $\phi$ and 
$K^{*0}$ mesons with $\lambda = 1$ and $\lambda = 3$ in Au+Au collisions at $\sqrt{s_{\rm NN}} = 200$ GeV}
\vspace{1.5mm}
\begin{tabular}{|c|c|cc|}
\hline
Centrality & ~Meson~ &\multicolumn{2}{c|}{$\rho_{00}^{\text{th}}$}  \\
\cline{3-4}
(\%) & & $\lambda=1$ & $\lambda=3$  \\
\hline
 ~ 0--15~~& $\phi$   &~ 0.333335~ & 0.333350~   \\
 ~15--30~~& $\phi$   &~ 0.333345~ & 0.333438~   \\
 ~30--60~~& $\phi$   &~ 0.333380~ & 0.333755~   \\
 ~ 0--15~~& $K^{*0}$ &~ 0.333335~ & 0.333347~   \\
 ~15--30~~& $K^{*0}$ &~ 0.333344~ & 0.333430~   \\
 ~30--60~~& $K^{*0}$ &~ 0.333376~ & 0.333714~   \\
\hline
\end{tabular}
\label{tab-phi-cent}
\end{table}

\subsection{Quantization-axis dependence}

We have also calculated the alignment assuming spin quantization along the $x$-axis. In this case, one has to calculate the alignment $A_1$, see Eq.~(\ref{eq:Ai}). Our calculations of $A_1$ give again positive alignment, which does not agree with the experimental result that shows that $A_1$ is negative. 


\section{Summary}
\label{sec:summary}

We have performed the analysis of spin alignment of $\phi$ and $K^{*0}$ vector mesons within a thermal model for heavy-ion collisions at the top RHIC energies. A common spin local equilibrium was assumed for $\Lambda$ hyperons and vector mesons. Our main findings can be summarized as follows:

\begin{enumerate}
    \item The model predicts, in contrast to many other physical analyzes~\cite{Liang:2004xn, Yang:2017sdk, Sheng:2019kmk, Sheng:2022wsy, Kumar:2023ghs}, a positive alignment and monotonic increase of $\rho_{00}$ with transverse momentum and centrality.
    
    \item The mass ordering is observed, with the heavier $\phi$ meson exhibiting a slightly larger deviation of $\rho_{00}$ from $1/3$ compared to $K^{*0}$.
    
    \item Quantitatively, the model with $\lambda = 1$ produces deviations $\rho_{00} - 1/3$ at the $10^{-4}$ level, which are below the experimental values. Increasing $\lambda$ to 3 enhances the deviations to the $10^{-3}$ level (for the largest considered values of transverse momentum), improving the alignment, but still cannot explain quantitatively the data.
\end{enumerate}

The fact that changes in the scale parameter $\lambda$ simultaneously affect the magnitude of the longitudinal polarization of $\Lambda$ and the spin alignment of the vector mesons suggests that there is a correlation between these two observables that deserves a more detailed study in the future. 

\section*{Acknowledgements}
Soham Banerjee and A.J. gratefully acknowledge the Department of Atomic Energy (DAE), India, for financial support. Samapan Bhadury acknowledges the support from Anusandhan National Research Foundation (ANRF), India through National Post Doctoral Fellowship, File No. PDF/2025/004233. W.F. and R.R. were supported in part by the National Science Centre (NCN), Poland, Grant No. 2022/47/B/ST2/01372. Samapan Bhadury, W.F. and R.R. would like to acknowledge the kind hospitality of the National Institute of Science Education and Research (NISER), Jatni, India, where this analysis was initiated.

\bibliographystyle{apsrev4-1}
\bibliography{apssamp}

\end{document}